\documentclass[aps,graphicx,showpacs]{revtex4}
\usepackage{graphicx}

\draft
\def\<{\langle}
\def\>{\rangle}

\begin{document}

\title{Controlled order rearrangement encryption for quantum key
distribution}
\author{Fu-Guo Deng$^{1,2}$, and G. L. Long$^{1,2,3,4}$\thanks{%
Corresponding author:gllong@tsinghua.edu.cn}}
\address{$^{1}$Department of Physics, Tsinghua University, Beijing 100084,
P. R. China\\
$^{2}$Key Laboratory For Quantum Information and Measurements,
Beijing 100084, P. R. China\\
$^{3}$Center for Atomic and Molecular NanoSciences, Tsinghua
University, Beijing 100084, P. R. China\\
$^{4}$Institute of Theoretical Physics, Chinese Academy of Sciences, Beijing 100080,
P. R. China}
\date{\today}

\begin{abstract}
A novel technique is devised to perform orthogonal state quantum
key distribution. In this scheme, entangled parts of a quantum
information carrier are sent from Alice to Bob through two
quantum channels. However before the transmission, the orders of
the quantum information carrier in one channel is reordered so
that Eve can not steal useful information. At the receiver's end,
the order of the quantum information carrier is restored. The
order rearrangement operation in both parties is controlled by a
prior shared control key which is used repeatedly in a quantum
key distribution session.
\end{abstract}

 \maketitle

 A complex telecommunication
system connects any place at any time with pervasive intrusion in
the world today. It is trivial to observe that information
security is a fundamental issue today. The task of cryptography is
to make secret messages intelligible only for the two legitimate
parties of the secret communication, Alice and Bob, and unreadable
for other unauthorized users such as Eve. To this end, Alice and
Bob have to encrypt their secret messages using a suitable
encryption scheme. Thus far, the only proven secure crypto-system
is the one-time-pad scheme, in which the secret key is as long as
the messages \cite{Vernam,Shannon}. The security of the
transmission of the secret messages using one-time-pad depends
ultimately on the key privacy. The security of key distribution is
the most important part in secret communication. Quantum key
distribution (QKD), an approach exploiting quantum mechanics
principles for secret communication, provides a secure way for
transmitting the key. A lot of attention has been focused on QKD
\cite{BB84,Ekert,BBM,B92,hwang1,GV,HIGM,Phoenix,KI,Brub,CabelloL,GLG,XLG,LL,LCA}
since the BB84 QKD protocol \cite{BB84}, and experimental studies
on QKD
have been developing very fast in the last two decades \cite%
{Bennett,MT,PT,Muller,Brendel,Tittel,Buttler20,Ribordy,Funk}.

The security of QKD lies on the fundamental difference between
classical and quantum information. Classical information can be
copied freely and imperceptibley. However quantum information
cannot be cloned \cite{Wootters}. Any measurement will disturb
the quantum state unless the quantum state is the measuring
device's eigenstate. For a quantum state, Eve has only one chance
to choose the right measuring device to avoid capture. The
security of QKD protocols lies either on randomness, e.g. in BB84
\cite{BB84} and similar protocols \cite{Ekert,BBM,B92,hwang1} or
the non-locality nature of quantum systems as in refs
\cite{GV,KI,CabelloL,LL}.

Non-locality is pertinent to quantum system only. Here a quantum state is
split into two parts: e.g. the two parts of a photon wave packet \cite{GV,KI}%
, or two correlated particles \cite{CabelloL,LL}. We call them as
quantum information carriers (QIC). In non-locality based QKD
protocols, orthogonal quantum states are used. Security is
assured by not allowing Eve to acquire both parts simultaneously.
These protocols can be understood in Fig.1 in ref \cite{GV},
which is similar to Fig.1. Alice and Bob's sites are secure, and
the transmission lines are insecure. Alice produces EPR pairs
randomly in one of entangled states. She then sends out the two
parts of a QIC simultaneously through two paths to Bob. The lower
QIC part is delayed first at Alice's site, and the upper QIC part
is sent away without delay. At Bob's site, the upper part has a
delay and the lower part has not, so the two parts of the QIC
arrive at the detector simultaneously and are measured. These
protocols use orthogonal states and have full efficiency; and all
the particles transmitted can be used to generate secret keys. In
Goldenberg-Vaidman scheme \cite{GV}, the time delay for the two
correlated parts is usually longer than the transmission time
between the practical distance so that the first QIC part has
already reached Bob's secure field when the second part starts to
run into the insecure line. To assure the security, Alice has to
send the QIC in random timing. Koashi-Imoto protocol \cite{KI}
uses an asymmetric interferometer instead of a symmetric
interferometer, and the random timing can be dropped and the
time-delay can be reduced. However, two factors make this delay
cannot be too short. First there exists a detection time window
for Eve in these protocols. If Eve acquires the same apparatus as
Bob, which is usually assumed in security analysis, she can take
an intercept-resend attack. She first intercepts a batch of lower
QIC parts which reaches her first and stores them for a while, at
the same time she sends fake lower QIC parts to Bob. After delay
time t, Eve begins to receive corresponding upper QIC parts.
Combining the QIC parts she keeps, she can make collective
measurement and know the correct key completely. Afterwards she
can send the correct messages to Bob. It is only in this delay
time that Eve makes errors about the secret keys. Afterwards,
Alice and Bob can only notice a backward shift of Bob's key string
relative to Alice's. Secondly, Eve can take advantages of noises
in noisy channel to gradually make up this shift. If the delay
time is not long, Eve can achieves this fairly quickly. Similar
issues exist for other protocols that use orthogonal states.

In this paper, we present a controlled order rearrangement
encryption (CORE) technique for QKD with the state of a correlated
system, such as entangled photon pairs. Here Alice rearranges the
order of correlated particles and send them to Bob, and Bob then
restores the order of the particles and recover the correct
correlation and makes the right measurement. This is done in a
controlled manner by the repeated use of a short control key as
has been used in the modified BB84 scheme\cite{hwang1}, where
Alice and Bob synchronize their measuring devices by repeatedly
using a prior shared control key.

To present our idea clearly, we use EPR pairs as the QIC in the
rest of this paper for the sake of simplicity. The procedure and
conclusions to other QIC's are very much the same with little or
without any modification. An EPR pair can be in one of the four
Bell states:
\begin{eqnarray}
\left\vert \psi ^{-}\right\rangle &=\frac{1}{\sqrt{2}}(\left\vert
0\right\rangle _{A}\left\vert 1\right\rangle) _{B}-\left\vert
1\right\rangle _{A}\left\vert 0\right\rangle _{B}, \;\;\;\;
\left\vert \psi ^{+}\right\rangle& =\frac{1}{\sqrt{2}}(\left\vert
0\right\rangle _{A}\left\vert 1\right\rangle _{B}+\left\vert
1\right\rangle
_{A}\left\vert 0\right\rangle _{B})\nonumber\\
\left\vert \phi ^{-}\right\rangle& =\frac{1}{\sqrt{2}}(\left\vert 0\right\rangle
_{A}\left\vert 0\right\rangle _{B}-\left\vert 1\right\rangle _{A}\left\vert
1\right\rangle _{B}), \;\;\;\;
 \left\vert \phi ^{+}\right\rangle
&=\frac{1}{\sqrt{2}}(\left\vert 0\right\rangle _{A}\left\vert 0\right\rangle
_{B}+\left\vert 1\right\rangle _{A}\left\vert 1\right\rangle _{B})  \label{EPR4}
\end{eqnarray}%
where subscripts A and B indicate the two correlated photons in each pair. They can
represent 00, 01, 10 and 11 respectively. As in Fig.1, CORE technique uses two
channels. Alice uses a modulator to prepare her EPR pairs randomly in the four
Bell-basis states, and then sends out them in equal time intervals to Bob. Before
these EPR pairs enter the insecure transmission lines, their orders are rearranged by
a controlled-order-rearrangement-encryption system. After they arrive at Bob's site,
they are de-arranged in Bob's site that undoes the effect of order rearrangement by
Alice and then are measured. Fig.2 shows the main idea of CORE by a specific example.
Here are four choices of CORE operations. The CORE is done for every four EPR pairs.
The upper QIC parts are transmitted according to their temporal ordering. A control
key is used to rearrange the order of particles in the lower channel. If the value of
control key is 00, operation E$_{0}$ is applied and the order of the 4 EPR pairs are
not changed as shown in Fig.2 (a), which is implemented in the device in Fig.2 (b)
with switches 1, 2 and 3 in positions (up, up, down) for all the 4 EPR pairs. When
control key is 01, E$_{1}$ is performed, and it is done by putting the 3 switches
into position (down, up, down), (up, down, up), (up, down, down), (up, down, up) for
the four particles respectively. Similar combinations can be written explicitly for
E$_{2}$ and E$_{3}$. After the order rearrangement, two particles that travel
simultaneously through the two channels have complicated relations: they are
correlated if the CORE operation is E$_{0}$, and they are anti-correlated if the
other 3 CORE operations are used. At Bob's sites, we just exchange upper and lower
part of Alice's CORE apparatus, and the CORE operation performed by Alice will be
undone. After measurement, the information is read out. It is interest to emphasize
that the measurement here is orthogonal basis measurement. The outcome is
determinative, and is exactly the same as Alice has prepared.

To prevent Eve stealing the information of the encryption ordering operation, one
needs an evening process to make the time interval between different batches of QIC's
travel at equal time intervals.

Let's look at the security of CORE. We suppose that Eve has complete
knowledge of Bob's measuring device except the control key. As she does not
know which CORE operation Alice and Bob are doing each time, she can only
guess randomly from the four CORE operations. Thus she has only 25\% chance
to choose the right CORE operation for every 4 EPR pairs. When Alice uses
the wrong CORE operation, the two photons she measures will be
anti-correlated, say the A particle from the first EPR pair and the B
particle from the second EPR pair is mistreated by Eve as an EPR pair, then
the density operator will be
\begin{equation}
\rho _{A_{1}B_{2}}=\overline{\rho }_{A_{1}}\otimes \overline{\rho }%
_{B_{2}}=\left(
\begin{array}{cccc}
1/4 & 0 & 0 & 0 \\
0 & 1/4 & 0 & 0 \\
0 & 0 & 1/4 & 0 \\
0 & 0 & 0 & 1/4%
\end{array}%
\right)   \label{matrix}
\end{equation}%
where $\overline{\rho }_{A_{1}}=Tr_{B_{1}}(\rho _{A_{1}B_{1}})$ and $%
\overline{\rho }_{B_{2}}=Tr_{A_{2}}(\rho _{A_{2}B_{2}})$ are the reduced
density matrices of particle A$_{1}$ and particle B$_{2}$, respectively.
When $\rho _{A_{1}B_{2}}$ is measured in Bell-basis, the result can be any
of the 4 Bell-basis states with 25\% probability each. Thus Eve will
introduce 3/4 $\times $ 3/4=56.25\% error rate in the results. Alice and Bob
can detect Eve easily by checking a sufficiently large subset of results
randomly chosen.

Surely, Eve can even take Bell inequality measurement on the photons, but it is
useless for decrypting  the control key. Let us choose
two directions $\overrightarrow{a}(a_{x},a_{y},a_{z})$ and $\overrightarrow{b%
}(b_{x},b_{y},b_{z})$ as the directions of measurement of Alice
and Bob respectively. Then the correlation operator can be written
as follow.
\begin{eqnarray}
\widehat{E} &=&\widehat{\sigma }\cdot \overrightarrow{a}\otimes \widehat{%
\sigma }\cdot \overrightarrow{b} \\
&=&\left[
\begin{array}{cccc}
a_{z}b_{z} & a_{z}(b_{x}-ib_{y}) & (a_{x}-ia_{y})b_{z} &
(a_{x}-ia_{y})(b_{x}-ib_{y}) \\
a_{z}(b_{x}+ib_{y}) & -a_{z}b_{z} & (a_{x}-ia_{y})(b_{x}+ib_{y}) &
-(a_{x}-ia_{y})b_{z} \\
(a_{x}+ia_{y})b_{z} & (a_{x}+ia_{y})(b_{x}-ib_{y}) & -a_{z}b_{z} &
-a_{z}(b_{x}-ib_{y}) \\
(a_{x}+ia_{y})(b_{x}+ib_{y}) & -(a_{x}+ia_{y})b_{z} & -a_{z}(b_{x}+ib_{y}) &
a_{z}b_{z}%
\end{array}%
\right]
\end{eqnarray}%
 where $\widehat{\sigma }$ is the Pauli operator,
$\sigma _{x}=\left(
\begin{array}{cc}
0 & 1 \\
1 & 0%
\end{array}%
\right) $, $\sigma _{y}=\left(
\begin{array}{cc}
0 & -i \\
i & 0%
\end{array}%
\right) $, $\sigma _{z}=\left(
\begin{array}{cc}
1 & 0 \\
0 & -1%
\end{array}%
\right) $. The expectation value $\left\langle E(\overrightarrow{a},%
\overrightarrow{b})\right\rangle _{\psi }=\left\langle \psi \right\vert
\widehat{\sigma }\cdot \overrightarrow{a}\otimes \widehat{\sigma }\cdot
\overrightarrow{b}\left\vert \psi \right\rangle $ in state $\left\vert \psi
\right\rangle $ is different for different Bell states. They are $%
-\left( a_{x}b_{x}+a_{y}b_{y}+a_{z}b_{z}\right) $, $%
a_{x}b_{x}+a_{y}b_{y}-a_{z}b_{z}$, $-a_{x}b_{x}+a_{y}b_{y}+a_{z}b_{z}$ and $%
a_{x}b_{x}-a_{y}b_{y}+a_{z}b_{z}$ in Bell states $\left\vert \psi
^{-}\right\rangle $, $\left\vert \psi ^{+}\right\rangle $, $\left\vert \phi
^{-}\right\rangle $ and $\left\vert \phi ^{-}\right\rangle $ respectively.
For product states $\left\vert 0\right\rangle \left\vert 0\right\rangle $, $%
\left\vert 0\right\rangle \left\vert 1\right\rangle $, $\left\vert
1\right\rangle \left\vert 0\right\rangle $ and $\left\vert 1\right\rangle
\left\vert 1\right\rangle $, the expected values are $a_{z}b_{z}$, $%
-a_{z}b_{z}$, $-a_{z}b_{z}$ and $a_{z}b_{z}$ respectively. If Eve takes Bell
inequality measurement on two uncorrelated  photons in the CORE repeatedly, she will
get 0 for a large number of measurements as the photons is randomly distributed in
the four product states. If Eve does takes two correlated photons, she will also get
 0 as the EPR pair takes the 4 Bell states with equal probability.
In fact, the violation of Bell inequality holds for two photons in a a fixed Bell
state. So Eve can get no information about the control key except for guessing it
randomly.

The control key is very important. Here we must emphasize that
unlike classical one-time-pad, the control key for CORE can be
quite short and be used repeatedly. This is surprising to a
conventional cryptographer because it is well known that a
one-time-pad key used twice will be of great danger. But it is a
different story when quantum mechanics comes into play. There are
fundamental differences in encryption between classical signal
and quantum state. In classical encryption, signal can be copied
freely and the copying process does not introduce error on the
final results. Eve can exploit this freedom to eavesdrop in
classical communication. In quantum physics, quantum states can
be measured without disturbing the system only if they are
eigenstates of the measuring operator. When Eve tries to measure
the states without knowing if she is using the eigenstate
operator, she will produce errors most of the time. The
no-cloning theorem\cite{Wootters} forbids Eve to produce multiple
copies to break the code by trial and error, and she has only one
chance for a quantum state! Different from classical
cryptography, Eve has no means to decipher the control key. She
can only guess the control key randomly. The
probability that Eve guesses the right control key is $(\frac{1}{4}%
)^{N_{k}}=(\frac{1}{2})^{2N_{k}}$, where 2N$_{k}$ is the number
of bits in the control key. When N$_{k}$=100, the probability is
practically zero.   The security of the repeated use of a short
control key has recently been proved\cite{hwang2}. It is shown
that QKD without public announcement of basis is secure against
both individual and coherent attacks. The proof could be applied
to this CORE scheme with some modification. More detailed study
needs be done to prove the security of the CORE scheme.

The control key is short. A few hundreds bit is enough for many
purposes. We can use a control key repeatedly in a single QKD
session. The control key can be produced in many ways. Any
sequence of secret numbers can be used as control key. They can
be produced beforehand when Alice and Bob are in contact. But the
preferred generating method is to produce them on-sites using the
same physical setup. Instead of simultaneously choosing the same
CORE operation, Alice and Bob choose their operation randomly.
They record their results and the CORE operation they use each
time. They have 25\% chance to choose identical CORE operations.
In these events, they should have identical results. After some
transmissions, they publish their CORE operations, and retain
those with identical CORE operations. They then perform
eavesdropping check. If the error rate is lower than a threshold,
they then conclude that their transmission is safe and then
continue to perform he follow-up post-processing such as error
correction and privacy amplification. Then they have a common
secret random numbers that can be used as the control key. After
the generation of the control key, Alice and Bob then switch to
the synchronized operation CORE, which is much more efficient.
Actually, the on-site generation of the control key is a
BB84-type protocol, and the intrinsic is low. However, we need
only to use this operation to generate a very short sequence key.
The time it takes is negligible compared to the main process of a
QKD process.

The CORE technique is not only suitable for EPR pairs, but also suitable for
other QIC's. The formalism used here can be directly translated into the
wave-packet case as in refs \cite{GV,KI}. In that case, the experimental
implementation will be easier. In addition, CORE can be performed in groups.
The control key can be used to control the CORE operation of a group of
units. For instance, instead of using 01 controls CORE operation of one unit
of QICs(4 EPR pairs) , one can use 01 to control the use of E$_1$ for more
units of QICs consecutively, say 4 units or 16 EPR pairs.

The specific example of CORE in this paper uses only four simple
permutation operations. The idea presented here can be easily
extended to build more complicated systems. The essence of CORE
technique is the repeated use of a control key to perform
classical encryption on the quantum system. The non-cloning
nature of quantum state ensures this technique viable.

Though genuine single photon source has been realized in
laboratory\cite{Schields,Yuan}, at present highly attenuated laser pulses is used to
approximate single photon source where only 1 out of 10 pulse contains a photon. The
detection efficiency is not 100\% either.  As with other QKD schemes,  we must perform
eavesdropping check, and if the error rate is less than threshold the results are
taken as the raw key. After quantum error correction and privacy amplification, the
raw key will be processed into ideal secret keys.

To summarize, CORE technique can be used to perform  secure key distribution with
present technology in a controlled and efficient manner. It is worth mentioning, the
discovery of security of repeated use of a short control key\cite{hwang1} is very
important. It enables quantum key distribution in a more efficient way.

This work is supported by the National Fundamental Research Program Grant
No. 001CB309308, China National Natural Science Foundation Grant No.
60073009, the Hang-Tian Science Fund, and the Excellent Young University
Teachers' Fund of Education Ministry of China.

\vskip 2cm

%\begin{figure}[tbp]
\begin{figure}
\begin{center}
\includegraphics[width=12cm,height=5cm]{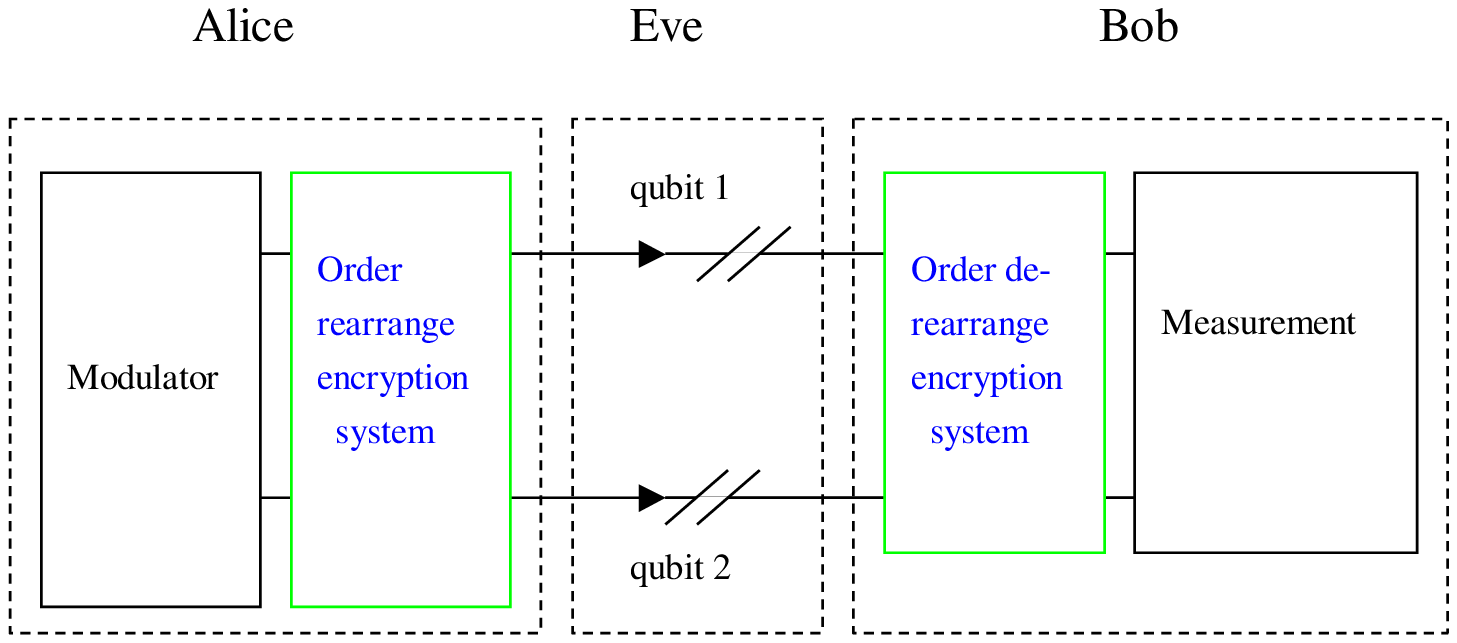}
\caption{Illustration of a typical QKD system with order rearranging encryption
system.}
\end{center}
\end{figure}

\begin{figure}
\begin{center}
\includegraphics[width=7cm,height=5cm]{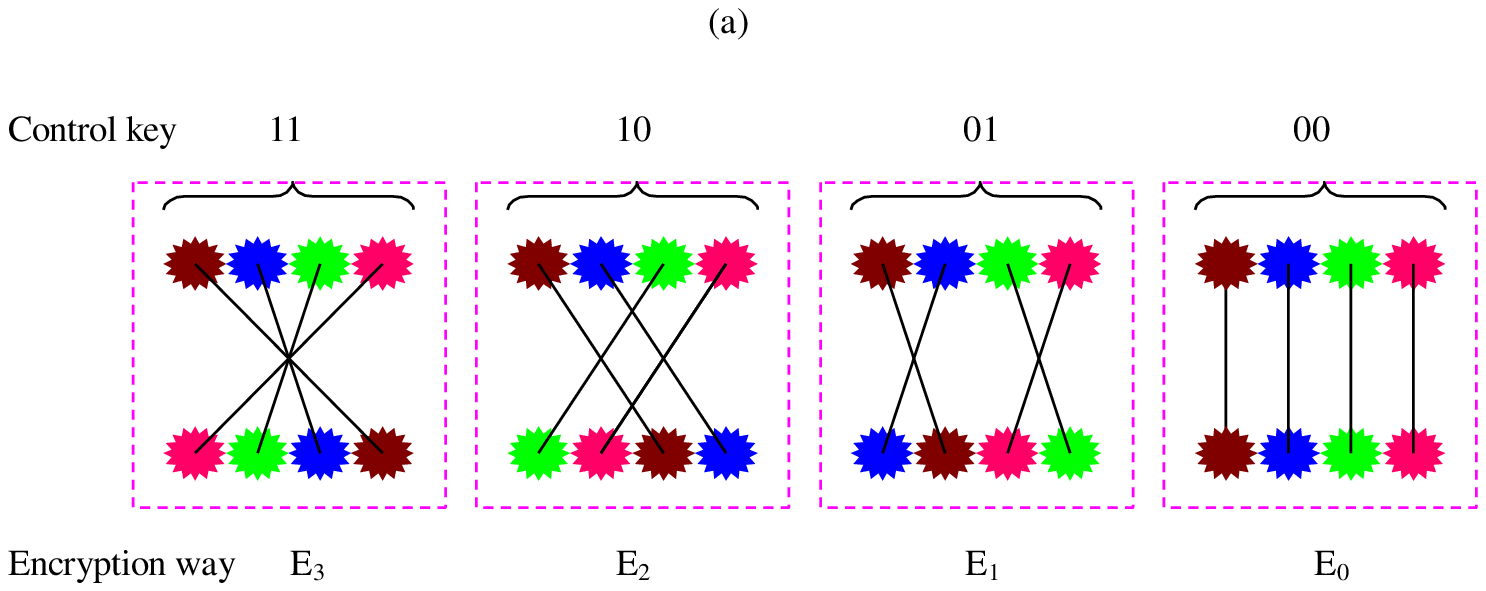}
\includegraphics[width=7cm,height=5cm]{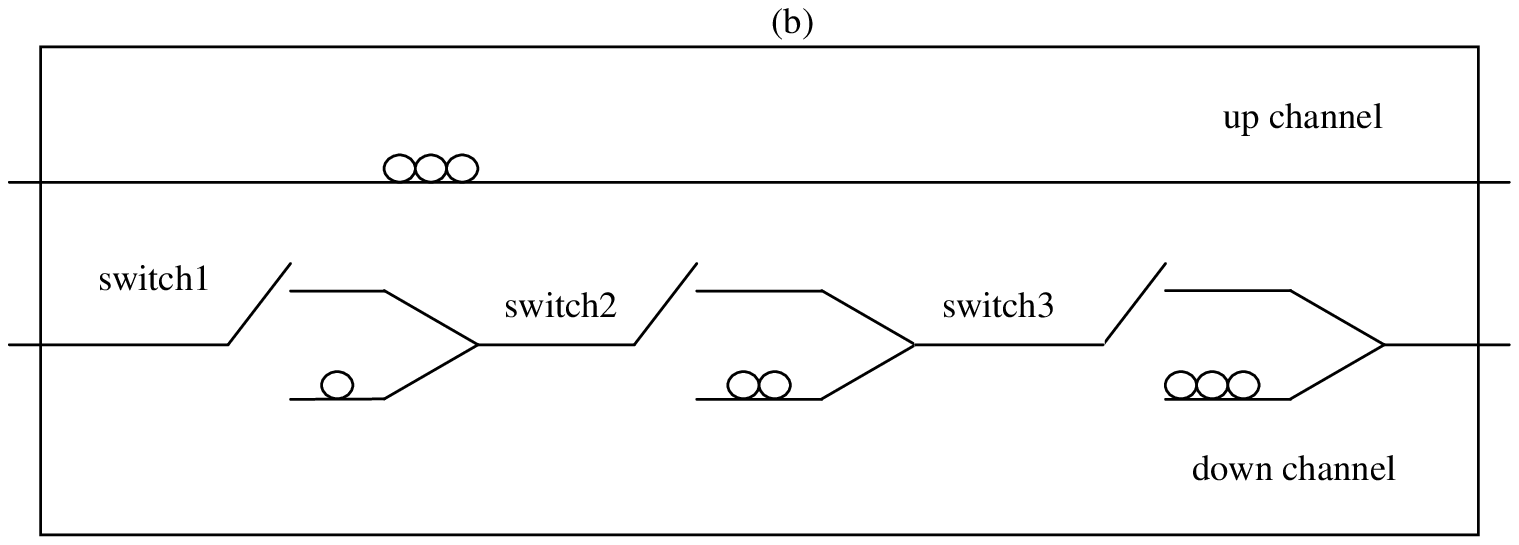}
\caption{A specific example of CORE with EPR pairs. (a) Four different CORE
operations.  (b) Device to perform CORE operations. The loop represents a time delay
of a fixed interval. }
\end{center}
\end{figure}


\begin{references}
\bibitem{Vernam} G. Vernam, J. Amer. Inst. Elec. Eng. 55, 109 (1926).

\bibitem{Shannon} C. E. Shannon, Bell System Technical Journal Vol. 28, 656
(1949).

\bibitem{BB84} C. H. Bennett, and G. Brassad, Proc. IEEE Int.Conf. on
Computers, Systems and Signal Processing, Bangalore, India (IEEE, New York,
1984), PP.175-179.

\bibitem{Ekert} A.Ekert, Phys. Rev. Lett. 67, 661 (1991).

\bibitem{BBM} C.H.Bennett, G.Brassard and N.D.Mermin, Phys.Rev.Lett. 68, 557
(1992).

\bibitem{B92} C.H.Bennett, Phys.Rev.Lett, 68, 3121 (1992).

\bibitem{hwang1}W. Y. Hwang, I. G. Koh and Y. D. Han, {\it Phys.
Lett.} {\bf A244}, 489 (1998).

\bibitem{GV} L. Goldenberg and L. Vaidman, Phys. Rev. Lett. 75, 1239(1995).

\bibitem{HIGM} B. Huttner, N. Imoto, N. Gisin and T. Mor,  Phys. Rev. A 51,
1863 (1995).

\bibitem{Phoenix} S. J. D. Phoenix et al., J. Modern Optics 42, 1155 (1995).

\bibitem{KI} M. Koashi and N. Imoto, Phys. Rev. Lett. 79, 2383 (1997).

\bibitem{Brub} D. Brub$\beta $, Phys. Rev. Lett. 81, 3013 (1998).

\bibitem{CabelloL} A. Cabello, Phys. Rev. Lett. 85, 5635 (2000).

\bibitem{GLG} G. P. Guo, R. Li, G. C. Guo, Phys. Rev. A 64, 042301 (2001).

\bibitem{XLG} P. Xue, C. F. Li and G. C. Guo, Phys. Rev. A 65,  022317
(2002).

\bibitem{LL} G. L. Long and X. S. Liu, Phys. Rev. A 65, 032302 (2002).

\bibitem{LCA} H. -k. Lo, H. F. Chan, M. Ardhali, e-print arXiv:
quant-ph/0011056.



\bibitem{Bennett} C.H. Bennett et al., J. Cryptography 5, 3 (1992).

\bibitem{MT} C. Marand, P.D. Townsend, Optics Lett. 20, 1695 (1995).

\bibitem{PT} S.J.D. Phoenix, P.D. Townsend, Contemporary Phys. 36, 165
(1995).

\bibitem{Muller} A. Muller et al., Appl. Phys. Lett. 70, 793 (1997).

\bibitem{Brendel} J.Brendel et al., Phys. Rev. Lett. 82, 2594 (1999).

\bibitem{Tittel} W.Tittel et al., Phys. Rev. A 59, 4150 (1999).

\bibitem{Buttler20} W.T Buttler et al., Phys. Rev. Lett. 84, 5652 (2000).

\bibitem{Ribordy} G. Ribordy et al., Phys. Rev. A 63, 012309 (2001).

\bibitem{Funk} A.C. Funk, M.G. Raymer, Phys. Rev. A 65, 042307 (2002).

\bibitem{Wootters} W. K. Wootters, W. H. Zurek, Nature 299, 802 (1982).

\bibitem{hwang2} W.-Y. Hwang, X. Wang, K. Matsumoto, J. Kim, H.-W. Lee,
Phys. Rev. A67, 012302(2003).

\bibitem{Schields} A. J. Schields et al, App. Phys. Lett. 76, 3673 (2000).

\bibitem{Yuan} Zhiliang Yuan et al, Science 295, 102 (2002).
\end{references}
\end{document}